\documentclass[twocolumn,showpacs,aps,prb,floatfix]{revtex4}
\usepackage{graphicx}
\usepackage{amssymb}
\usepackage{longtable}
\usepackage{threeparttable}
\usepackage{amsmath}
\usepackage{amsfonts}
\usepackage{color}
\usepackage{soul}

\begin{document}
\title
{Direct observation of the Mott gap in Sr$_2$IrO$_4$ with a scanning tunneling microscope
}
\author{ John Nichols$^1$, Noah Bray-Ali$^{2,3}$, Gang Cao$^1$ and Kwok-Wai Ng$^1$}

\affiliation
{
$^1$Center for Advanced Materials, Department of Physics and Astronomy, University of Kentucky, Lexington, Kentucky 40506-0055, USA\\
$^2$Joint Quantum Institute, University of Maryland, College Park, Maryland 20742-4111, USA\\
$^3$National Institute of Standards and Technology, Gaithersburg, Maryland 20899, USA\\
}

\begin{abstract}

The single-layer Mott insulator Sr$_2$IrO$_4$ was studied using a scanning tunneling microscope.  This measurement technique is unique due to the transport properties of this Mott insulator allowing tunneling measurements to be performed, even at cryogenic temperatures.  We obtained high-resolution images of the sample surface and the differential tunneling conductance at different cryogenic temperatures.  The differential conductance is a direct measurement of the local electronic density of states which provided an insulating gap consistent with optical conductivity, angle resolved photoemission spectroscopy and resonant inelastic x-ray scattering (RIXS) experiments.  The observed widths of these features is broader than predicted by the Slater approximation and narrower than predicted by dynamical mean field theory.  Additionally, the observed density of states due to magnetic fluctuations is found in the derivative of the differential conductance and is consistent with results from Raman scattering and RIXS.  At low temperatures, additional low-energy features were observed, suggesting a change in the dispersion of the collective magnetic excitations, which is consistent with the magnetic susceptibility.  

\end{abstract}

\pacs{73.23.Hk, 71.30.+h, 71.70.Ej, 71.27.+a}

\maketitle


\section{Introduction}

Recently, there has been a surge of renewed interest in 5\textit{d} transition metal oxides, such as the iridates. The electrons in these compounds are more spatially extended than in their 3\textit{d} and 4\textit{d} counterparts, resulting in a smaller Coulomb interaction (U) and a larger bandwidth (W). This should make 5\textit{d} transition-metal oxides metallic; however, it is believed that  the large spin-orbit coupling (SOC) in the 5\textit{d} materials drives some of them insulating.  The prototype of this spin-orbit-driven insulating behavior is the single-layered iridate, Sr$_2$Ir0$_4$.

Tunneling spectroscopy is the most direct method for probing a material's density of states, which for Mott insulators is not well understood and often has contributions from multiple excitations.  The interesting phenomenon with this material is that the SOC parameter is large enough to allow even a modest Coulomb interaction to open a Mott gap.  We use scanning tunneling spectroscopy (STS) to directly measure the Mott gap.  Since tunneling spectroscopy is mostly sensitive to the charge property of electrons, magnon excitations are rarely seen in tunneling experiments.  Interestingly, in the present study, the SOC in Sr$_2$IrO$_4$ allows an anomalous magnon feature to be seen.  

\begin{figure}[h!]
    \centering
    \graphicspath{{./Final/}}
    \includegraphics[width= 7 cm]{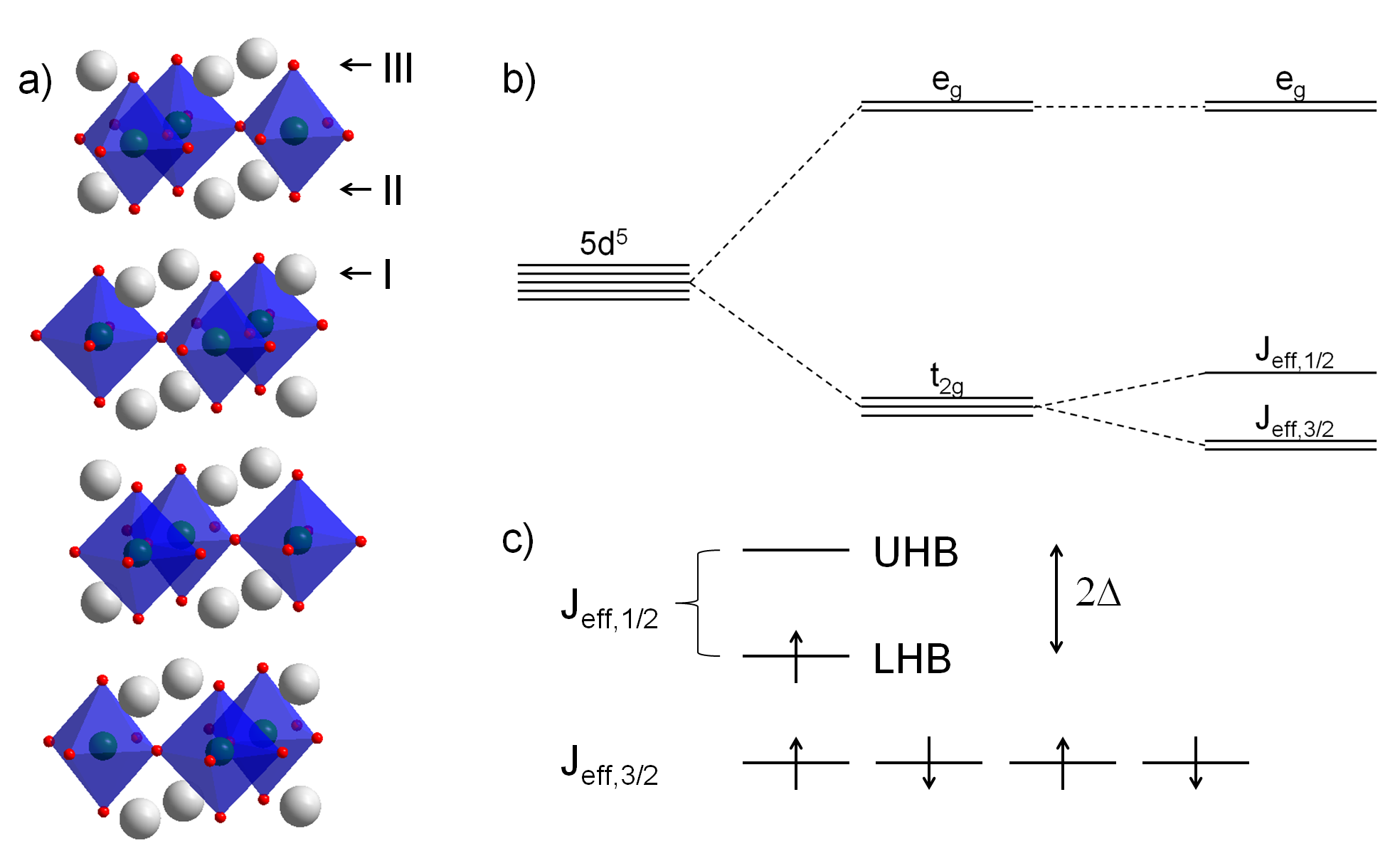}
    \caption{a) Crystal structure of Sr$_2$IrO$_4$.  The Sr, Ir and O atoms are white, green and red respectively.  The IrO$_6$ octahedra are colored blue. The three layers labeled I, II and III are labels for the layers in Fig. \ref{fig:image:618_627}.  b) Illustration of how the 5d orbital is split due to crystal field splitting and spin-orbit coupling.  c) Illustration of the effect of the Coulomb interaction on the J$_{\mathrm{eff,1/2}}$ band and the ground state configuration of the system.} 
    \label{fig:split_struc}
\end{figure}

Sr$_2$IrO$_4$ has a K$_2$NiF$_4$-like crystal structure as illustrated in Fig. \ref{fig:split_struc} a.  The IrO$_6$ octahedra undergo a rotation about the c-axis of approximately $11^\circ$ resulting in a reduced tetragonal crystal structure (space group I4$_1$/acd).  To the knowledge of the authors, the metal insulator transition in undoped Sr$_2$IrO$_4$ has not been observed, and this system has been shown to be insulating for temperatures up to 600 K. \cite{chikara:prb:09}  This material has five valence electrons to fill the 5\textit{d} orbital.  Crystal fields split this band into a lower-energy t$_{\mathrm{2g}}$ band and a higher-energy e$_\mathrm{g}$ band.  SOC then splits the t$_{\mathrm{2g}}$ band into a J$_{\mathrm{eff,3/2}}$ that is filled with four electrons, and the remaining electron half fills the more highly energetic J$_{\mathrm{eff,1/2}}$ state.  The effects of band splitting due to crystal fields and SOC are shown in Fig. \ref{fig:split_struc} b.  The modest Coulomb interaction in the J$_{\mathrm{eff,1/2}}$ band forces this material into an insulating state \cite{kim:prl:08} as shown in Fig. \ref{fig:split_struc} c.  

Remarkably, even though Sr$_2$IrO$_4$ has an insulating ground state, it maintains a finite, non-zero electrical conductance even at cryogenic temperatures.\cite{chikara:prb:09}  Tunneling experiments have been successfully performed on other Mott-insulating, transition-metal-oxide systems such as Ca$_3$Ru$_2$O$_7$ \cite{bautista:ssc:08} and Ca$_2$CuO$_2$Cl$_2$. \cite{ye:arxiv:12}  In this Rapid Communication, we report measurements of the atomic-scale lattice structure and electronic structure of single crystalline Sr$_2$IrO$_4$ using a scanning tunneling microscope (STM).

Measurements of the electronic structure of the spin-orbit-driven Mott insulator Sr$_2$IrO$_4$ that use other techniques have been previously reported.  SrIrO4 is an insulator that has an energy gap with a full width defined as 2$\Delta$.  Measurements of the excitation gap using resonant inelastic x-ray scattering (RIXS) and angle-resolved photoemission spectroscopy (ARPES) suggest values of 2$\Delta$ = 400 meV \cite{kim:prl:12} and 2$\Delta$ = 580 meV, \cite{dassau} respectively. Optical conductivity vanishes for photon energies below 2$\Delta$ $\sim$ 400 meV,\cite{moon:prb:09} suggesting a value for the insulating gap close to the RIXS value.  We measured the energy gap in the local density of states using STM.  This local probe of the electronic structure provides a complementary perspective to measurements with other techniques.

\section{Methods}

Our experiments were conducted with a home-built Pan-style STM\cite{pan:rsi:99} with a commercial controller.  The STM was oriented vertically inside a cryogenic probe, allowing for \textit{in-situ} sample cleaving and exchange.  The probe was evacuated to P $\sim10^{-7}$ Torr at room temperature, and then the pressure was lowered further by cryogenically cooling the entire experimental probe.  The samples were oriented such that the c-axis was parallel to the STM tip.  The crystals were prepared with the flux method.  The samples were cleaved \textit{in-situ} at T = 77 K by breaking off a small rod that was attached to the top of the sample.  The weakest bond is the Sr-O bond between an IrO$_6$ octahedra and an Sr atom in neighboring atomic layers.  This weak bond leaves the SrO layer to be exposed after cleaving.  Thus, the IrO$_2$ layer, which is responsible for this material's electrical conduction, lies underneath an insulating layer, and this limits our resolution to single atomic steps.

We interpret the electronic structure of Sr$_2$IrO$_4$ using the half-filled, one-band Hubbard model on the square lattice with on-site Coulomb interaction energy U = 2.0 eV and nearest-neighbor hopping amplitude t = 0.25 eV.\cite{jin:prb:09,wang:prl:11}

Below the Neel temperature $T_N=240$ K, canted antiferromagnetic order arises in the J$_\textrm{eff}=1/2$ Ir moments.\cite{Kim:science:09,Cao:prb:98}  Due to the rotation of the IrO$_6$ octahedra, the canted antiferromagnetic order leads to a bulk ferromagnetic moment. \cite{wang:prl:11}  Bulk magnetometry yields a staggered moment of $|\vec{S}_0| \approx 0.307$.\cite{wang:prl:11}  It has been reported that Sr$_2$IrO$_4$ has additional magnetic ordering at temperatures of $\sim$ 100 K and $\sim$ 25 K.\cite{ge:prb:11}

The lattice parameters for $\mathrm{Sr}_2\mathrm{IrO}_4$ are a = 5.499 \AA\hspace{3 pt}and c = 25.799 \AA . \cite{crawford:prb:94}  It is important to note that the tilt of the IrO$_6$ octahedra quadruples the c-axis lattice parameter.  Therefore, the distance between two adjacent layers of IrO$_6$ octahedra is roughly $c/4\sim$ 7 \AA.  This material is inhomogeneous \cite{korneta:prb:10} and cleaves along the SrO planes.

High-resolution topographic images of the sample surface were obtained using an electrochemically etched tungsten tip.  Tunneling current in topography-mode was I = 200 pA, sample bias voltage was V$_\textrm{s}$ = 300 mV, and the tip voltage was kept at virtual ground.  

STS was obtained using a lock-in amplifier to measure the differential tunneling conductance $dI/dV$ as a function of sample bias.  STS were obtained at both 4.2 K and 77 K.  For the 4.2 K (77 K) data, the sample bias was modulated by V$_{\mathrm{mod}}$ = 10 mV (4 mV) at a frequency of f$_{\mathrm{mod}}$ = 704.2 Hz (703.4 Hz).  We performed spatial averaging of the STS signal by using a blunt gold tip. \cite{kwapinski:ss:10}  

\begin{figure}[h!]
    \centering
    \graphicspath{{./Final/}} 
    \includegraphics[width=7 cm]{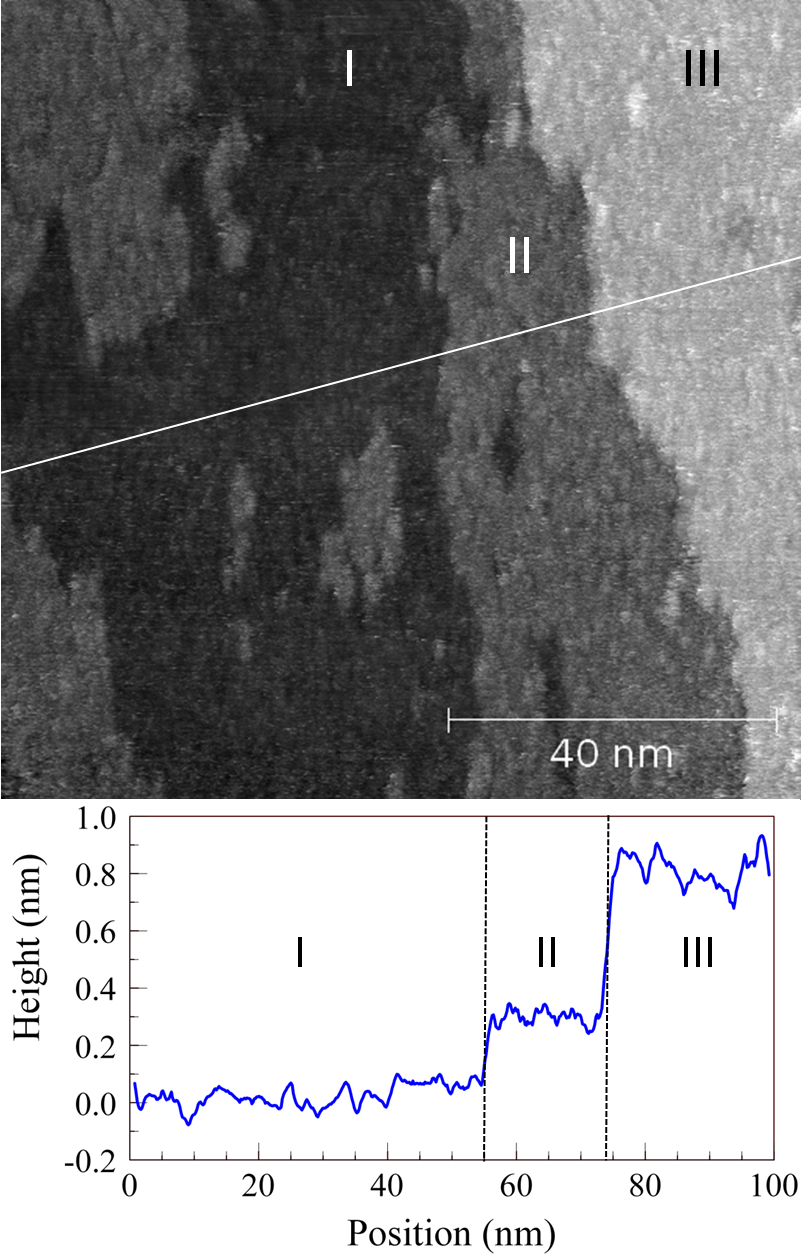}
    \caption{(top)  A constant-current STM image of the raw data taken with V$_{\mathrm{bias}}$ = 300 mV, I$_{\mathrm{set}}$ = 200 pA of 109 $\times$ 109 nm$^2$ area of $\mathrm{Sr}_2\mathrm{IrO}_4$.  Each layer labeled I, II and III is a different SrO layer, as indicated in Fig. \ref{fig:split_struc}. (bottom) A line profile represented as a white line in the image.} 
    \label{fig:image:618_627}
\end{figure}

\section{Results and Analysis}

\begin{figure}[h!]
    \centering
    \graphicspath{{./Final/}} 
    \includegraphics[width=7 cm]{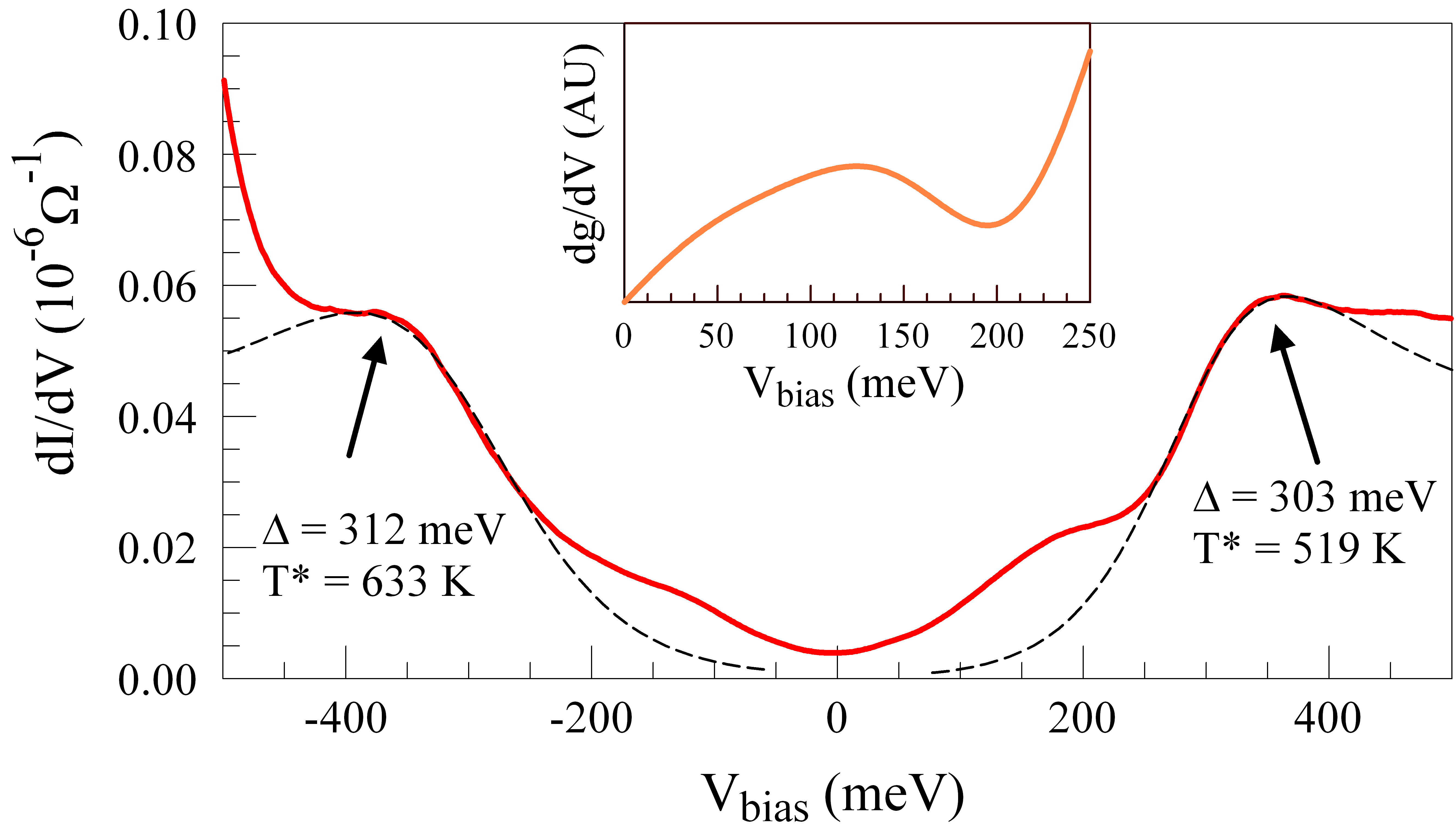}
    \caption{The red curve is the differential conductance at T = 77 K obtained from lock-in amplifier with V$_{\mathrm{mod}}$ = 4 mV,\;f$_{\mathrm{mod}}$ = 703.4 Hz.  The dashed lines are fits of the peaks to Eqn. \ref{didv_fit}.  The parameters to each fit are displayed.  Inset: The derivative of the even component of the differential conductance.} 
    \label{fig:didv_77k}
\end{figure}

\indent High-resolution topographic images of Sr$_2$IrO$_4$ are shown in Fig. \ref{fig:image:618_627}.  Step edges separate wide, atomically flat steps, labeled I, II, and III [top, field of view (109 $\times$ 109 nm$^2$ )].  Along the path indicated by the white line in the topographic image, the variation of height of the surface was measured (bottom) with atomic resolution in the direction parallel to the tip.  Note that STM measures only the electronic structure and not the atomic position.  No lattice structure was observed in the lateral direction due to the strong correlation of electrons in this system.  Since the distance between layers I and III is consistent with the separation of two adjacent octahedra layers, \citep{crawford:prb:94} and the STS in all three regions are identical (see below) suggests these layers have the same stoichiometry as indicated in Fig. \ref{fig:split_struc}.  STS were consistent between different cleaves.

The high-temperature tunnelling spectrum has a small density of states at the Fermi energy (Fig. \ref{fig:didv_77k}).  The spectrum has rough symmetry between positive and negative energy.  Away from the Fermi energy, the density of states rises and has a prominent peak.    A shoulder on the broad peak was observed on the side of the peak closer to the Fermi energy.  

The broad peaks in the differential tunneling conductance were fitted (dashed line, separate fits to peaks above and below Fermi energy) using the standard tunneling formalism \cite{bk:tinkham} and using the density of states of the half-filled, one-band Hubbard model on the square lattice \cite{wang:prl:11} within the Slater mean-field approximation \cite{bk:fradkin}:

\begin{equation}\label{didv_fit}
\frac{dI}{dV} \propto \int_{-\infty}^\infty dE \frac{|E|}{\sqrt{E^2-\Delta^2}} \frac{e^{(E+eV)/k_BT^*}}{k_BT^*(1+e^{(E+eV)/k_BT^*})^2}
\end{equation}

Here, $T^*$ and $\Delta$ are treated as fit parameters.  On physical grounds, we expect $T^* \approx 77$ K for these high-temperature STS.  Similarly, within the Slater approximation, the single-particle gap parameter is expected to be given by $2\Delta = \frac{4}{3}U|\vec{S}_0| \approx 800$ meV, where $U\approx 2.0$ eV and $|\vec{S}_0|=0.307$ are, respectively, the on-site Coulomb interaction and staggered magnetic moment.\cite{bk:fradkin,wang:prl:11}

From the fits to the STS, we infer the size of the insulating gap to be $2\Delta\approx 615$ meV.  Similarly, we infer the temperature parameter $T^* = 519$ K from fitting the peak at positive bias and $T^* = 633$ K from fitting the peak at negative bias.  The size of the gap is smaller than expected within the Slater approximation.  However, it is consistent
with the ARPES result, 2$\Delta$ = 580 meV.   The RIXS result, 2$\Delta$=400 meV, is smaller than the gap measured in STS.  Similarly, the onset energy for optical absorption 400 meV is
also smaller than the gap measured in STS. 

The broadening parameter obtained from fitting the data to Eqn. (1) is much larger than the temperature at which the STS were obtained.  We interpret this broadening as evidence of electron-electron correlation, which is not captured by the Slater mean-field approximation.  The peak in the density of states is narrower than expected from results obtained within the dynamical mean-field approximation to the one-band, half-filled Hubbard model on the square lattice. \cite{wang:prb:09}

Surprisingly, inside the Mott gap additional features that we interpret as inelastic loss features due to a single mangnon are observed. \cite{tsui:prl:71}  Since these features should be symmetric about the Fermi energy, the asymmetry of the Mott gap is removed by fitting the data in Fig. \ref{fig:didv_77k} (not shown), and defining the function g as the even component of this fit.  The derivative of g for the high-temperature differential tunneling spectrum in the vicinity of the low-energy shoulder is shown in the inset of Fig. \ref{fig:didv_77k}.  There is a peak at 125 meV, which is the energy associated with the single magnon. \cite{tsui:prl:71}  RIXS measurements of the magnon dispersion relation are consistent with this interpretation: the magnon dispersion flattens around $120$ meV creating a large increase in the density of states. \cite{kim:prl:12}  Similarly, the two-magnon absorption feature in Raman scattering \cite{ cetin:arxiv:12} is consistent with our interpretation:  the two-magnon feature in Raman scattering occurs at $240$ meV, roughly twice the energy of the single-magnon feature in STS.  

\begin{figure}[h!]
    \centering
    \graphicspath{{./Final/}} 
    \includegraphics[width=7 cm]{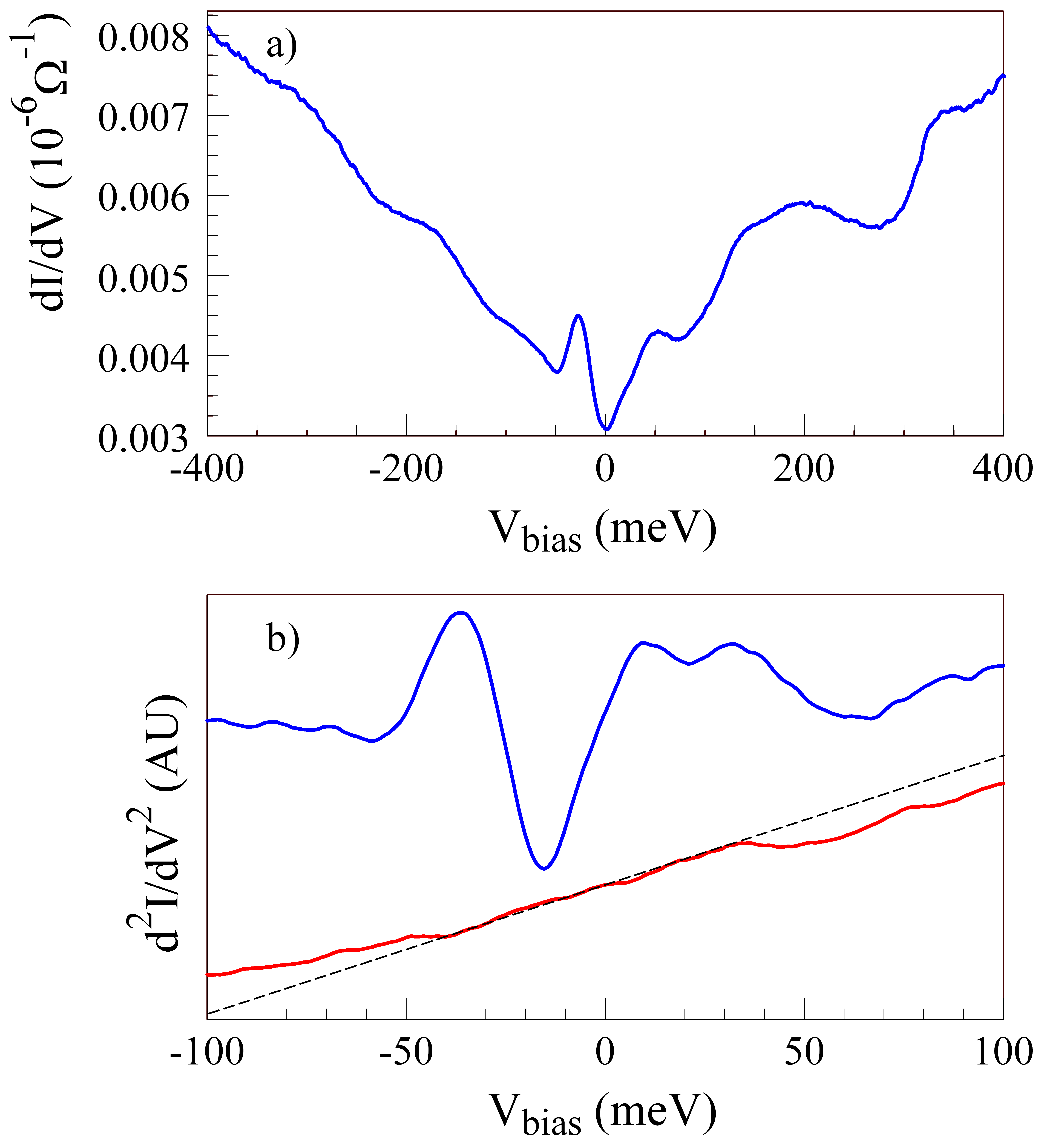}
    \caption{a) The differential conductance at T = 4.2 K obtained from lock-in amplifier with V$_{\mathrm{mod}}$ = 10 mV,\;f$_{\mathrm{mod}}$ = 704.2 Hz.  b) The derivative of the differential conductance at T = 77 K (red) and T = 4.2 K (blue).  The dashed line is a guide for the eye.} 
    \label{fig:didv_4k}
\end{figure}

The low-temperature tunneling spectrum (4.2 K) is shown in Fig. \ref{fig:didv_4k} a.   Suppression of the density of states at the Fermi energy is clearly visible.   Away from the Fermi energy, the density of states rises, just as it does at high temperature.  However, in addition to features due to the insulating gap and single magnon also observed at high temperature, there are clear rearrangements of spectral weight and a general decrease in the particle-hole symmetry of the tunneling spectrum at low temperature.  In particular, near the Fermi energy, an additional shoulder appears which is sharper than the prominent features seen in the high-temperature STS.  The low-temperature spectra was also fit to Eqn. \ref{didv_fit}, and a similar value for 2$\Delta$ was obtained; however, the parameters of these fits are very uncertain and are not presented.

The derivative of the low-temperature tunneling spectrum is shown in Fig. \ref{fig:didv_4k} b in red along with the high-temperature spectrum (blue) and a guide to the eye (dashed).  There is a narrow peak in the low-temperature data at approximately $35$ meV, which we interpret as  an inelastic loss feature due to excitation of a collective excitation of the low-temperature magnetic order which is consistent with the additional magnetic ordering recently observed in this material. \cite{ge:prb:11}    Bulk magnetometry and muon spin relaxation show a change in magnetic order at low temperature, with an onset at roughly 20 K.  The appearance of a new inelastic loss feature in the STS at low temperature is consistent with a change in magnetic order, and suggests that the density of states for magnetic excitations at low temperature has an additional peak at $35$ meV compared to the high-temperature density of states for magnetic excitations.

\section{Conclusions}

\indent We have examined the properties of Sr$_2$IrO$_4$ with an STM.  We have shown that its freshly cleaved surfaces are atomically flat with steps that are single atomic layers.  We have measured the Mott gap with tunneling spectroscopy to be $2\Delta = 0.615$ eV at 77 K.  This result is comparable to similar results from ARPES, RIXS and optical conductivity measurements.  We have modeled our results with the single-band Hubbard model and the Slater (dynamical mean-field theory) approximation, which predict features that are narrower (broader) than observed.  We have also observed a single-magnon excitation at an energy of 0.125 eV, which is consistent with magnons observed through Raman scattering and RIXS.  Also, at lower temperatures, we have observed additional low-energy features at roughly $\pm$ 35 meV, likely due to additional magnetic ordering. 

\section{Acknowledgements}

This research was supported by NSF grants DMR-0800367, DMR-0856234 and EPS-0814194.  Noah Bray-Ali acknowledges support from the National Research Council Postdoctoral Research Associateship Program.\\


\begin{thebibliography}{10}

\bibitem{chikara:prb:09}
S.~Chikara, O.~Korneta, W.~P. Crummett, L.~E. DeLong, P.~Schlottmann, and
  G.~Cao.
\newblock {\em Phys. Rev. B}, 80:140407, Oct 2009.

\bibitem{kim:prl:08}
B.~J. Kim, Hosub Jin, S.~J. Moon, J.-Y. Kim, B.-G. Park, C.~S. Leem, Jaejun Yu,
  T.~W. Noh, C.~Kim, S.-J. Oh, J.-H. Park, V.~Durairaj, G.~Cao, and
  E.~Rotenberg.
\newblock {\em Phys. Rev. Lett.}, 101:076402, Aug 2008.

\bibitem{bautista:ssc:08}
A.~Bautista, V.~Durairaj, S.~Chikara, G.~Cao, K.-W. Ng, and A.K. Gupta.
\newblock {\em Solid State Communications}, 148:240, 2008.

\bibitem{ye:arxiv:12}
Cun Ye, Peng Cai, Runze Yu, Xiaodong Zhou, Wei Ruan, Qingqing Liu, Changqing
  Jin, and Yayu Wang.
\newblock {\em arXiv:1201.0342v1}, 2012.

\bibitem{kim:prl:12}
Jungho Kim, D.~Casa, M.~H. Upton, T.~Gog, Young-June Kim, J.~F. Mitchell,
  M.~van Veenendaal, M.~Daghofer, J.~van~den Brink, G.~Khaliullin, and B.~J.
  Kim.
\newblock {\em Phys. Rev. Lett.}, 108:177003, Apr 2012.

\bibitem{dassau}
Q.~Wang, Y.~Cao, J.~A. Waugh, T.~F. Qi, O.~B. Korneta, G.~Cao, and D.~S.
  Dessau.
\newblock {\em submitted to Phys. Rev Lett., 2012}.

\bibitem{moon:prb:09}
S.~J. Moon, Hosub Jin, W.~S. Choi, J.~S. Lee, S.~S.~A. Seo, J.~Yu, G.~Cao,
  T.~W. Noh, and Y.~S. Lee.
\newblock {\em Phys. Rev. B}, 80:195110, Nov 2009.

\bibitem{pan:rsi:99}
S.~H. Pan, E.~W. Hudson, and J.~C. Davis.
\newblock {\em Review of Scientific Instruments}, 70:1459, 1999.

\bibitem{jin:prb:09}
Hosub Jin, Hogyun Jeong, Taisuke Ozaki, and Jaejun Yu.
\newblock {\em Phys. Rev. B}, 80:075112, Aug 2009.

\bibitem{wang:prl:11}
Fa~Wang and T.~Senthil.
\newblock {\em Phys. Rev. Lett.}, 106:136402, Mar 2011.

\bibitem{Kim:science:09}
B.~J. Kim, H.~Ohsumi, T.~Komesu, S.~Sakai, T.~Morita, H.~Takagi, and T.~Arima.
\newblock {\em Science}, 323(5919):1329--1332, 2009.

\bibitem{Cao:prb:98}
G.~Cao, J.~Bolivar, S.~McCall, J.~E. Crow, and R.~P. Guertin.
\newblock {\em Phys. Rev. B}, 57:R11039--R11042, May 1998.

\bibitem{ge:prb:11}
M.~Ge, T.~F. Qi, O.~B. Korneta, D.~E. De~Long, P.~Schlottmann, W.~P. Crummett,
  and G.~Cao.
\newblock {\em Phys. Rev. B}, 84:100402, Sep 2011.

\bibitem{crawford:prb:94}
M.~K. Crawford, M.~A. Subramanian, R.~L. Harlow, J.~A. Fernandez-Baca, Z.~R.
  Wang, and D.~C. Johnston.
\newblock {\em Phys. Rev. B}, 49:9198--9201, Apr 1994.

\bibitem{korneta:prb:10}
O.~B. Korneta, Tongfei Qi, S.~Chikara, S.~Parkin, L.~E. De~Long,
  P.~Schlottmann, and G.~Cao.
\newblock {\em Phys. Rev. B}, 82:115117, Sep 2010.

\bibitem{kwapinski:ss:10}
T.~Kwapi\'{n}ski and M.~Ja{\l}ochowski.
\newblock {\em Surface Science}, 604:1752, 2010.

\bibitem{bk:tinkham}
Michael Tinkham.
\newblock {\em Introduction to Superconductivity}.
\newblock Robert I. Krieger Publishing Company, 1980.

\bibitem{bk:fradkin}
Eduardo Fradkin.
\newblock {\em Field Theories of Condenses Matter Systems}.
\newblock Addision-Wesley Publishing: Redwood City, CA, 1991.

\bibitem{wang:prb:09}
Xin Wang, Emanuel Gull, Luca de' Medici, Massimo Capone, and Andrew~J. Millis.
\newblock {\em Phys. Rev. B}, 80:045101, Jul 2009.

\bibitem{tsui:prl:71}
D.~C. Tsui, R.~E. Dietz, and L.~R. Walker.
\newblock {\em Phys. Rev. Lett.}, 27:1729--1732, Dec 1971.

\bibitem{cetin:arxiv:12}
Mehmet~Fatih Cetin, Peter Lemmens, Vladimir Gnezdilov, Dirk Wulferding, Dirk
  Menzel, Tomohiro Takayama, Kei Ohashi, and Hidenori Takagi.
\newblock {\em arXiv:1201.3841v1 [cond-mat.str-el]}.

\end{thebibliography}
\end{document}